\documentclass[%
 reprint,
superscriptaddress,
 amsmath,amssymb,
 aps,
]{revtex4-1}

\usepackage{graphicx}
\usepackage{dcolumn}
\usepackage{bm}
\usepackage{xfrac}
\usepackage{comment}
\usepackage{mathtools}
\usepackage{physics}
\usepackage{makecell}
\newcommand*{\kb}{k_{\rm{B}}}

\begin{document}

\preprint{APS/123-QED}

\title{Non-equilibrium thermodynamics in a single-molecule quantum system
}

\author{E. Pyurbeeva}
\email{e.d.pyurbeeva@qmul.ac.uk}
 \affiliation{School of Physical and Chemical Sciences, Queen Mary University of London, Mile End Road, London E1 4NS, UK}
\author{J.O. Thomas}
\affiliation{Department of Materials, University of Oxford, Parks Road, Oxford, OX1 3PH, UK
}
\author{J.A. Mol}
\email{j.mol@qmul.ac.uk}
\affiliation{School of Physical and Chemical Sciences, Queen Mary University of London, Mile End Road, London E1 4NS, UK}%

\date{\today}
\begin{abstract}

Thermodynamic probes can be used to deduce microscopic internal dynamics of nanoscale quantum systems. Several direct entropy measurement protocols based on charge transport measurements have been proposed and experimentally applied to single-electron devices. To date, these methods have relied on (quasi-)equilibrium conditions between the nanoscale quantum system and its environment, which constitutes only a small subset of the experimental conditions available. In this paper, we establish a thermodynamic analysis method based on stochastic thermodynamics, that is valid far from equilibrium conditions,  is applicable to a broad range of single-electron devices and allows us to find the difference in entropy between the charge states of the nanodevice, as well as a characteristic of any selection rules governing electron transfers. We apply this non-equilibrium entropy measurement protocol to a single-molecule device in which the internal dynamics can be described by a two-site Hubbard model.  
\end{abstract}
\maketitle
\section{Introduction}
Correlated electronic states in nanoscale systems hold great promise for use in quantum technologies. Several direct entropy measurement protocols \cite{Hartman2018, Kleeorin2019, Sela2019, Pyurbeeva2020b} have been proposed to illuminate the microscopic dynamics of such states in solid-state devices, and are approaching practical applications in twisted bilayer graphene \cite{Saito2021, Rozen2021}, single-molecule devices \cite{Pyurbeeva2021, Gehring2021} or systems expected to host exotic quasiparticles \cite{Han2021, Child2021, Child2021a} . These newly developed experimental techniques are extremely powerful in probing quantum systems under quasi-static thermodynamic equilibrium conditions, but their underlying theoretical framework rapidly breaks down outside the linear response regime\cite{Pyurbeeva2022}.

The restriction to equilibrium or quasi-equilibrium conditions is a common theme in both thermodynamics and quantum transport. Despite the significant work done in the past thirty years in the field of non-equilibrium thermodynamics \cite{Seifert2012}, the true non-equilibrium realm beyond the linear response remains a ``dark zone'', where thermoelectric effects are usually treated phenomenologically, through the rate equation. However, the results that do exist typically apply to small systems with significant fluctuations \cite{Jarzynski1997, Crooks1999, Jarzynski2011}. Conveniently, this includes single-electron nanodevices \cite{pekola2015, Pekola2018}, which experience significant changes in both energy and particle number with every electron passing through them. Thus, applying non-equilibrium thermodynamic results to nanodevices can potentially offer a thermodynamic analysis of all the experimental data, rather than a small quasistatic subset. 

A related benefit of developing approaches that can treat systems out of equilibrium is that the introduction of a new energy scale, originating from the bias voltage, can allow access to higher-lying energy levels that have a negligible population in quasi-equilibrium conditions, and thus cannot be revealed in conventional entropy measurements.   

This paper offers an example of taking a stochastic thermodynamic approach to nanoscale charge transport. We apply the general non-equilibrium fluctuation relation \cite{Seifert2005, Schmiedl2007, Seifert2008} to a single-electron nanodevice, and propose a new method for measuring entropy and exploring electron-transfer selection rules in highly non-equilibrium systems (\emph{i.e.} at bias voltages much larger than any thermal fluctuation). We then test this method on experimental data of a single-molecule device, showing that the entropy changes between charge states agree with the energy level structure previously found from fitting the device stability diagram to the Hubbard dimer model.\cite{Thomas2021} Finally, we demonstrate that information on electron localisation in the dimer can be extracted from the tunneling current. 

\section{Theoretical development of the method}
\subsection{Non-equilibrium fluctuation theorem}
\begin{figure}
\includegraphics[width=\linewidth]{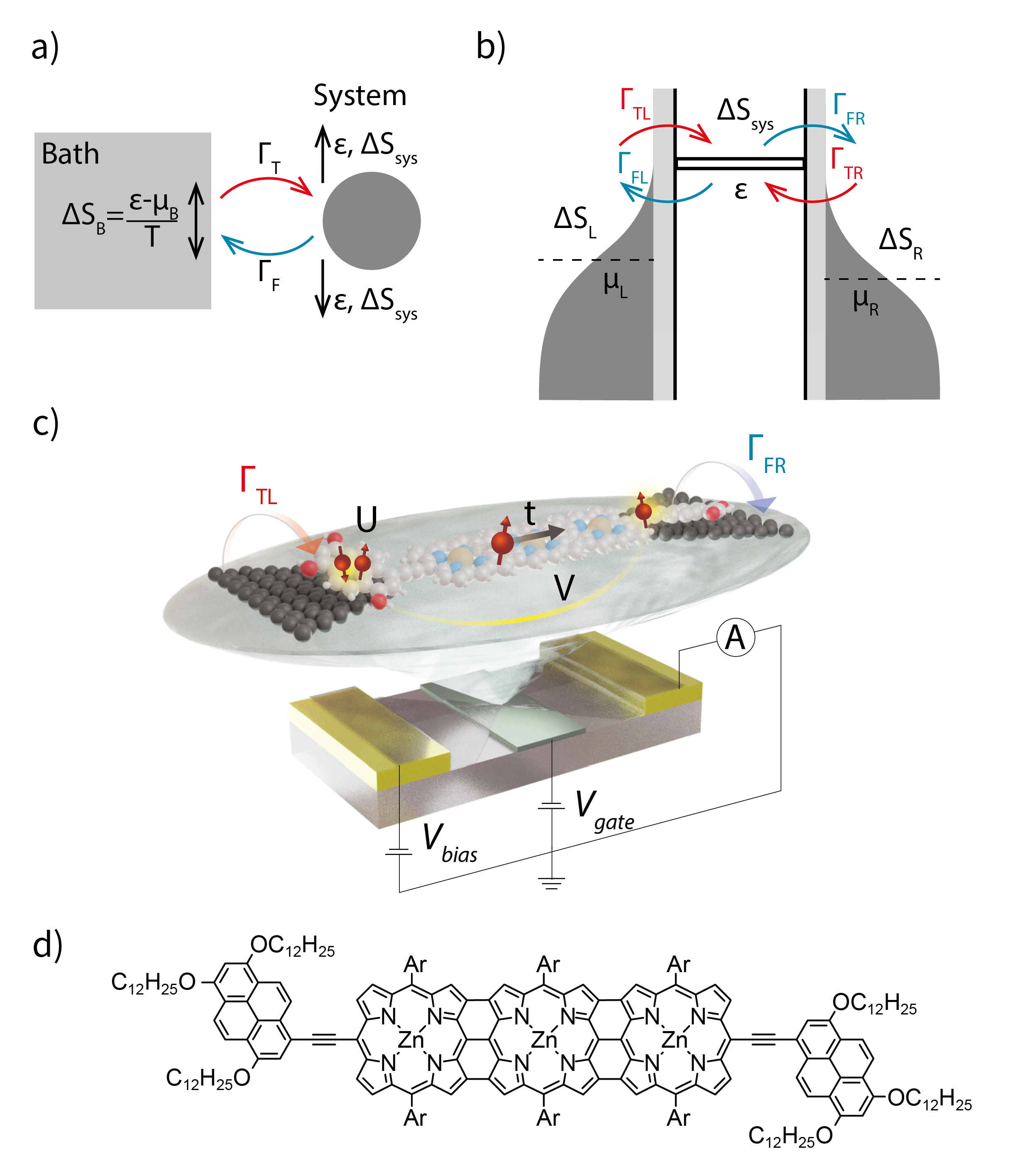}
\caption{a) A general view of the system exchanging particles with a thermal bath. The entropy changes in both the system and the bath are indicated.  b) An energy diagram of a typical charge transport measurement -- a transport energy level $\varepsilon$, corresponding to a single charge-state transition ($N \rightarrow N+1$) exchanges electrons with two electron baths at chemical potentials $\mu_L$ and $\mu_R$. The particle exchange with either bath is associated with a change of its entropy by $\Delta S_{L/R}$ and a change of system entropy by $\Delta S_{sys}$.  c) Architecture of a molecular device discussed in this work\cite{Thomas2021}: a molecule bridges the nanometre-sized electroburnt gap in a graphene ribbon\cite{Prins2011, Lau2014, Sadeghi2015, Pyurbeeva2021a}.  The graphene is patterned into a bowtie shape, in which the nanogap is made, with the sides forming the source and drain electrodes, and overlaps two gold electrodes. A back gate is used to shift the molecular energy levels. The Hubbard parameters, $U$ (intra-site potential energy), $V$ (inter-site potential energy) and $t$ (kinetic energy) are indicated schematically on the molecular structure. d) Molecular structure of the edge-fused porphyrin trimer molecule used in the experimental data\cite{Thomas2021}.}
\label{fig1}
\end{figure}
We consider the entropy of a few-electron system weakly coupled to an electron reservoir -- a thermal bath. Electrons can tunnel to and from the system with tunnel rates $\Gamma_T$ and $\Gamma_F$, respectively, as shown in Figure 1a. As an electron tunnels from the reservoir to the system, the entropy of the \emph{universe}, which is the sum of the entropy of the system and the entropy of the reservoir, changes by $+\Delta S$. Equally, as an electron tunnels from the system to the reservoir, the entropy of the universe changes by $-\Delta S$. Since the ratio of the tunnel rates to and from the system is equal to ratio of probabilities of the system fluctuating, we can apply the general non-equilibrium fluctuation theorem\cite{Seifert2005, Schmiedl2007, Seifert2008, pekola2015, Pekola2018}:
\begin{equation}
\label{eq-theorem}
    \frac{\Gamma_T}{\Gamma_F} = \frac{\Gamma(+\Delta S)}{\Gamma(-\Delta S)} = e^{\Delta S/k_B}.
\end{equation}
The main conceptual difference between this approach and the previous entropy measurement methods in nanodevices \cite{Hartman2018, Kleeorin2019, Sela2019, Pyurbeeva2020b, Pyurbeeva2021, Pyurbeeva2022}, which has to be emphasised, is that $\Delta S$ in Equation \ref{eq-theorem} is not the difference in entropy between the two charge states of the devices, but the change of the total entropy of the universe associated with the hopping process, equal to: $\Delta S = \left(\varepsilon - \mu\right)/T +\Delta S_{sys}$
where the first term corresponds to the change in entropy of the reservoir, given by the single-particle energy of the system $\varepsilon$, the chemical potential $\mu$ and temperature $T$ of the reservoir. The second term denotes the entropy difference between the two charge states of the system involved in the transition, the value used in previous conventions  \cite{Hartman2018, Kleeorin2019, Sela2019, Pyurbeeva2020b, Pyurbeeva2021}. 

We note that the average occupation of the system can be derived from the non-equilibrium fluctuation theorem and is consistent with that previously found from the Gibbs distribution and thermodynamic considerations \cite{Pyurbeeva2020b} (see Appendix \ref{proof-sec}). Moreover, the approach can be further generalised to include multiple reservoirs to enable application to the experimental single-electron transistor platform, as illustrated in Figure \ref{fig1}b, c. We give a detailed description of the molecular device below.

When multiple electronic states are involved in the charge transport through a  single-electron transistor and the change in the Fermi distribution $f(\varepsilon)$ is small on the scale of the level spacing, the tunnel rates to and from the system are given by:
\begin{equation}
\label{eq-rates}
\begin{cases}
    \Gamma_T &= \gamma d_{01}f(\varepsilon)\\
    \Gamma_F &= \gamma d_{10}\left(1-f(\varepsilon)\right)
\end{cases}
\end{equation}
Here and below the numeric subscripts indicate the excess number of electrons of the charge states between which the transfer occurs. The equation above concerns the $N\leftrightarrow N+1$ transfer, $\gamma$ is a geometric coupling coefficient given by the tunnel barrier between the system and the reservoir, $f(\varepsilon)$ is the Fermi-distribution in the reservoir, and $d_{01}$ and $d_{10}$ are the system-dependent coefficients for the 0 to 1 excess charge state and \emph{vice versa} transitions, given by the numbers and energies of the electronic states, and selection rules determined by Dyson coefficients. The transition probabilities $d_{01}$ and $d_{10}$ are generally not equal, and for the case of simple spin degenerate levels take integer values representing the degeneracy of the ``receiving'' state. 

Since $f(\varepsilon)/\left(1-f(\varepsilon)\right)=e^{\left(\varepsilon-\mu\right)/k_BT}$, Equations \ref{eq-theorem} and \ref{eq-rates} can be combined into an expression linking the entropy difference between the charge states of a few-electron system with the system-dependent coefficients $d_{01/10}$:
\begin{equation}
    \Delta S_{01} = k_B\ln\frac{d_{01}}{d_{10}}.
\end{equation}
where $\Delta S_{01}$ is the difference in entropy between the $N+1$ and $N$ charge states.

In the rest of this article we will demonstrate how this equation, derived from the general fluctuation theorem, can be used to directly measure the entropy of a molecular few-electron system far from equilibrium, as well as to find further information on the system's microscopic dynamics.

\subsection{Direct entropy measurement}
\begin{figure}
\includegraphics[width=\linewidth]{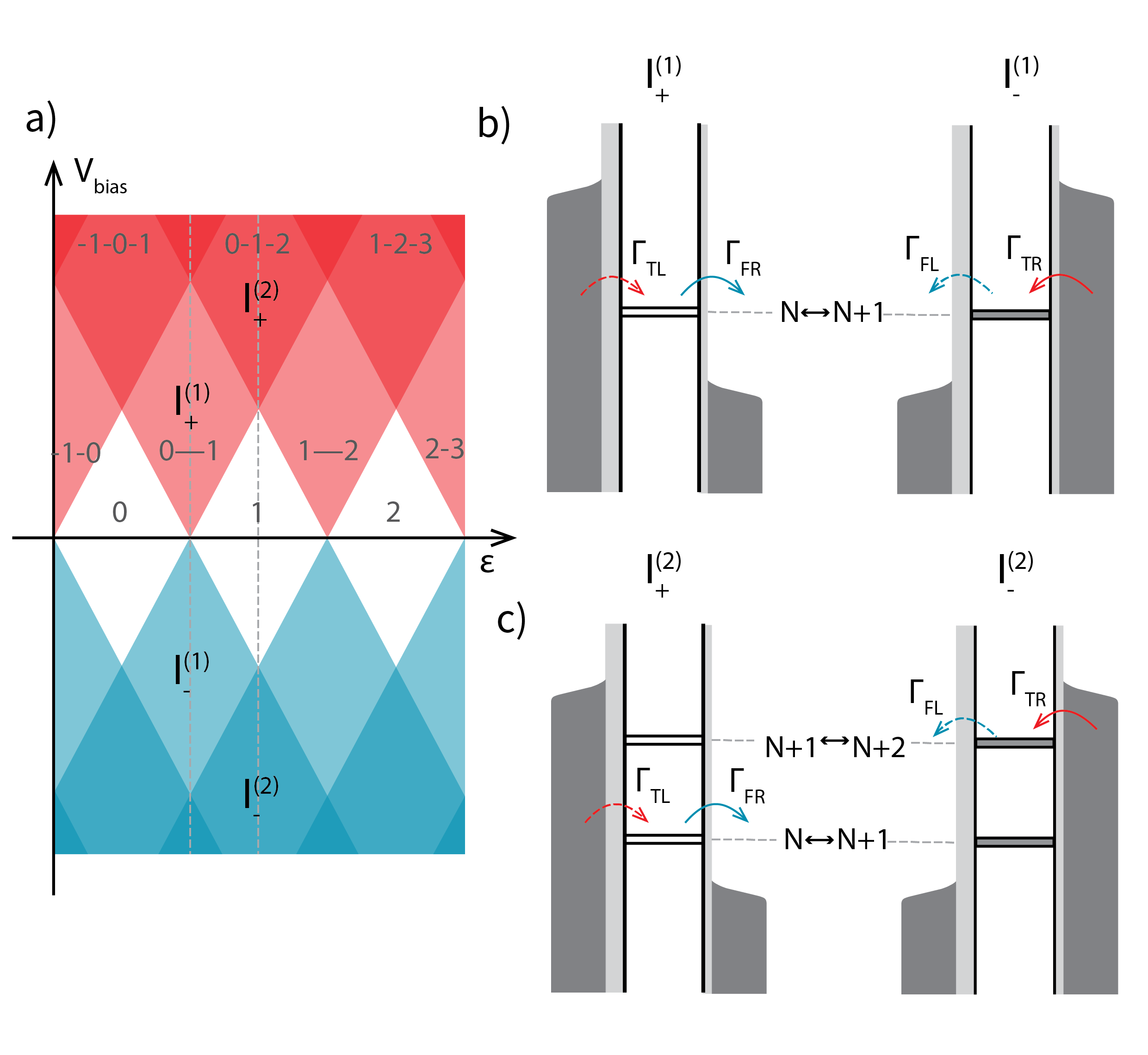}
\caption{a) A general stability diagram showing Coulomb diamonds -- areas of blocked current (white), areas with one energetically allowed charge-state transition (light colour) and those with two energetically allowed charge-state transitions (darker colour). b,c) Energy diagrams of the transitions and system population in the case of high tunnel coupling asymmetry and large bias in the different areas of the stability diagram -- with one (b) and two (c) energetically allowed charge-state transitions. The dashed arrows (through wide tunnel barriers) show the slowest electron-transfer processes. The fill of the transport energy level represents its population.} 
\label{fig2}
\end{figure}

We show that the entropy difference between the charge-states can be inferred from the current through a single-electron transistor if the following conditions are met: i) the applied bias voltage is sufficiently large such that the Fermi distribution of the leads does not vary significantly on the energy scale of the electronic excited states of the system (see Figure \ref{fig2}a and b); ii) the coupling to one lead is far greater than the coupling to the other lead ($\gamma_L\ll\gamma_R$ in Figure \ref{fig2}b); and iii) only transitions between two adjacent charge states are energetically accessible. The second condition is typically met in molecular single-electron transistors due to atomistic variations in the molecule-lead interactions\cite{Limburg2019}, and in solid-state devices couplings can be controlled using barrier gates. The first and third condition depend on the applied bias and gate voltage, as illustrated in Figure \ref{fig2}a.

When the above conditions are met, and we take the case: $\gamma_L\ll\gamma_R$, the non-equilibrium steady state current through a single energetically available charge-state transition is: $I^{(1)} = \Gamma_{TL} P_N - \Gamma_{FL} P_{N+1}$, where $P_{N/N+1}$ is the probability of finding the system in the $N/N+1$ charge state. At positive bias voltages this simplifies to $|I^{(1)}_+|=\Gamma_{TL}$, and at negative bias $|I^{(1)}_-|=\Gamma_{FL}$  as $P_N$ and $P_{N+1}$ are either 1 or 0 depending on the applied bias polarity.

From this, we find that the logarithm of the current ratio is a direct measure of the entropy difference between the $N$ and $N+1$ charge state of the system:
\begin{equation}
\label{eq-ratio}
        \ln\frac{I^{(1)}_+}{I^{(1)}_-}=\pm\Delta S_{01}/k_B,
\end{equation}
The freedom in sign represents the general case of an unknown direction of the asymmetry between the couplings $\gamma_L$ and $\gamma_R$.


\subsection{Two charge-state transitions}

Next, we consider a combination of bias and gate voltage such that two charge-state transitions are energetically allowed, as shown in Figure \ref{fig2}c. Since a total of three charge-states are involved in charge transfer, direct application of Eq. \ref{eq-ratio} is not applicable.

Again, we begin with the assumption of highly asymmetric tunneling barriers. In one bias direction the system occupies the 0 excess charge state the majority of the time, with the current proportional to the rate of the slowest occurring process (the rate-determining step), and thus the  transition probability $d_{01}$. Under the opposite bias, the system occupies the 2 excess charge state, occasionally switching to 1,  and the current is proportional to $d_{21}$.

The logarithm of the ratio of the currents is equal to:
\begin{equation}
    \ln\frac{I^{(2)}_+}{I^{(2)}_-}= \pm \ln \frac{d_{01}}{d_{21}}= \pm \ln \left[ \frac{d_{01}}{d_{10}} \frac{d_{10}}{d_{12}} \frac{d_{12}}{d_{21}} \right]
\end{equation}
Using Eq. \ref{eq-ratio}, and the final form of the above equation, we can express the logarithm of the ratio of the currents in the two hopping processes regime as:
\begin{equation}
\label{eq-two}
    \ln\frac{I^{(2)}_+}{I^{(2)}_-}=\pm \left(\Delta S_{01} + \Delta S_{12} + \sigma_{02}\right)/k_B
\end{equation}
where the first two terms on the right-hand side of the equation correspond to the entropy differences between consecutive charge states. The third term, $\sigma_{02} = k_B\ln d_{10}/d_{12}$ is a measure for the relative probability of positive and negative charge-state fluctuations starting at excess charge state 1 -- a parameter not previously considered.

\section{Method application}
\subsection{Experimental data analysis}
\begin{figure}
\centering
\includegraphics[width=0.5\textwidth]{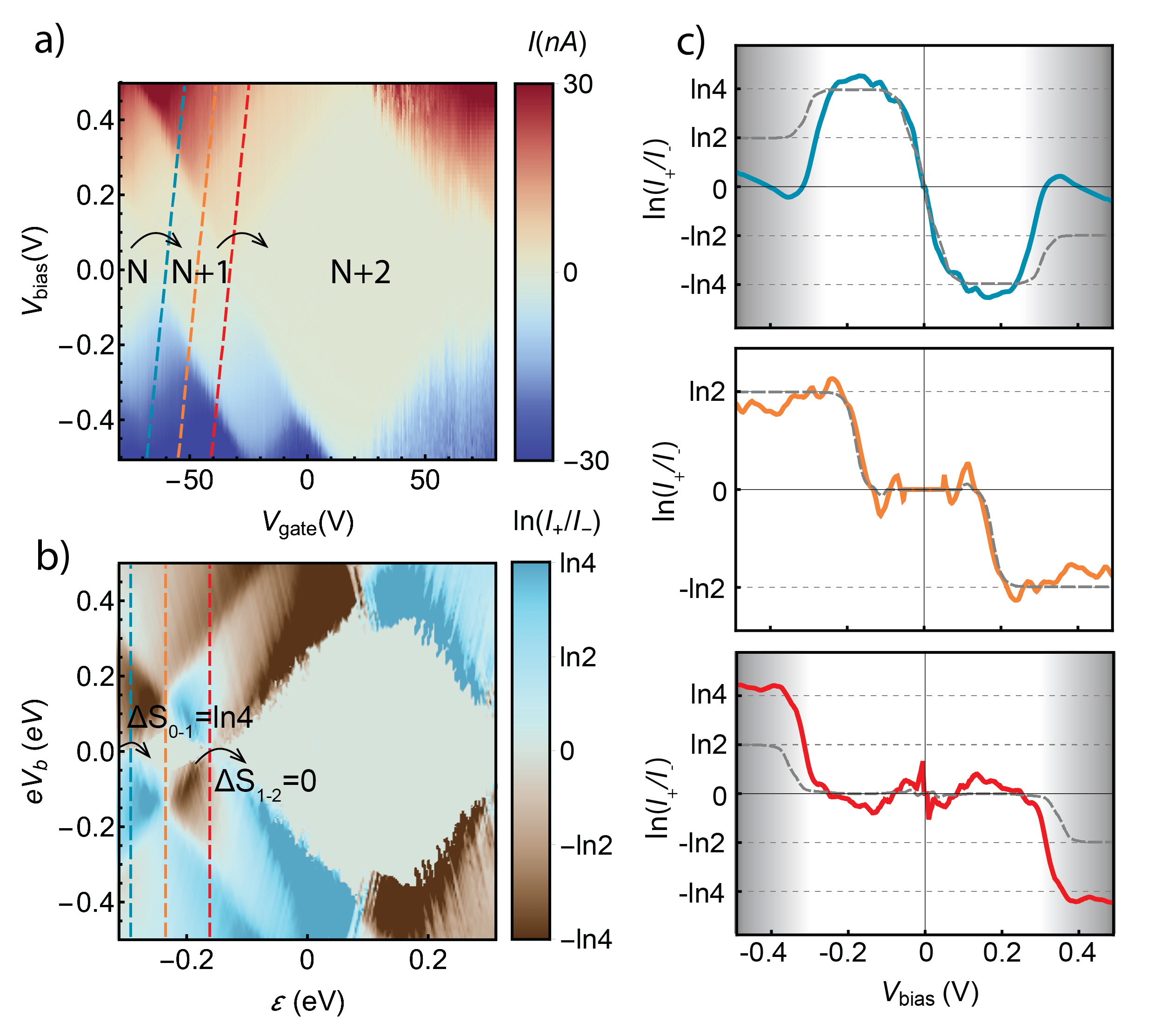}
\caption{a) Experimental current stability diagram of the porphyrin trimer single-molecule device, measured at 77K. Data from \cite{Thomas2021}. The charge-state assignment of the Coulomb diamonds is indicated. b) Map of the logarithm of the ratio of the positive and negative bias currents the same device, as a function of bias gap $eV_b$ and electrostatic contribution to the transport energy level, $\varepsilon$ (given by $V_{g}$, multiplied by gate lever arm of $4.8 \times 10^{-3}$. c) Linear cuts of the logarithmic map through the $N\rightarrow N+1$ transition resonance, $N-1/N/N+1$ degeneracy point and $N+1\rightarrow N+2$ transition resonance, shown in a) and b) in corresponding colours. Thick lines show the interpolated experimental data, while the dashed grey lines follow from the Hubbard model in \cite{Thomas2021}. The shading shows areas with significant deviation from the model due to vibrational effects (see Appendix \ref{vibrations-sec}).}
\label{fig3}
\end{figure}

\begin{table}
\caption{\label{table-states}The eigenstates and eigenvalues of the Hubbard Hamiltonian\cite{Fransson06} for the charge states involved in the experimental data \cite{Thomas2021}, assuming equal site energies. $\Phi_A=\ket{\uparrow \downarrow,0}$, $\Phi_B=\ket{0,\uparrow \downarrow}$, $\Phi_C=\ket{\uparrow, \downarrow}$, $\Phi_D=\ket{\downarrow, \uparrow}$. The coefficients $c_+$ and $c_-$ depend on the Hubbard parameters $U$, $V$ and $t$ (see Fig. \ref{fig1}c and Appendix \ref{dyson-sec}) and were found in \cite{Thomas2021} to be equal to 0.71 and 0.04 respectively. $C=\sqrt{(U-V)^2/4+4t^2}$}
\begin{ruledtabular}
\begin{tabular}{llll}
\makecell{Charge\\ state} & \makecell{State\\ label} & \makecell{Eigenstates of $H_{HB}$} & \makecell{Level energy}\\
\hline\\
$N$ & $S^{0}$ & $\ket{0,0}$ & $0$\\[5pt] \hline
\\
$N+1$ & $D_{+, \uparrow/\downarrow}^{1}$ & $(\ket{\uparrow, 0}+\ket{0,\uparrow})/\sqrt{2}$ &$t$\\[5pt]
&  & $(\ket{\downarrow, 0}+\ket{0,\downarrow})/\sqrt{2}$ &
\\[5pt]
 & $D_{-, \uparrow/\downarrow}^{1}$ & $(\ket{\uparrow, 0}-\ket{0,\uparrow})/\sqrt{2}$ &$-t$\\[5pt]
&  & $(\ket{\downarrow, 0}-\ket{0,\downarrow})/\sqrt{2}$ &
\\[5pt] \hline\\
$N+2$  & $S_{CS}^{2}$ & $(\ket{\uparrow \downarrow, 0}-\ket{0, \uparrow \downarrow})/\sqrt{2}$ &$U$ \\[5pt]
 & $T_{-1,0,1}^{2}$ & $(\ket{\uparrow, \downarrow }-\ket{\downarrow, \uparrow})/\sqrt{2}$ &$V$\\[5pt]
& & $\ket{\uparrow, \uparrow}$ &\\[5pt]
& & $\ket{\downarrow, \downarrow}$ &
\\[5pt]
& $S_{-}^{2}$ & $c_-(\Phi_A+\Phi_B)-c_+(\Phi_C-\Phi_D)$ &$\dfrac{U+V}{2}-C$
\\[5pt]
 & $S_{+}^{2}$ & $c_+(\Phi_A+\Phi_B)+c_-(\Phi_C+\Phi_D)$ &$\dfrac{U+V}{2}+C$\\[5pt]
\end{tabular}
\end{ruledtabular}
\end{table}
We test our analysis method on data from taken for a single-molecule device.\cite{Thomas2021} The molecule studied is a fused porphyrin trimer functionalised with two pyrene anchoring groups that $\pi$-stack onto graphene electrodes, the molecular structure is shown in Fig. \ref{fig1}d. The coupling, $\gamma$, that results from $\pi$-stacking is weak,\cite{Limburg18} leading to transport through the molecule being dominated by single-electron tunneling and Coulomb blockade (Fig. \ref{fig3}a).

The charge stability diagram features three charge states: $N$, $N+1$, and $N+2$ (Fig. \ref{fig3}a). The device displays asymmetric coupling to the electrodes ($\gamma_R > \gamma_L$), as is common in single-molecule devices, allowing for the fluctuation relation approach to be applied. A small asymmetry is present in the coupling between the bias voltage applied and the energy levels of the molecule, causing the skewedness of the Coulomb diamonds. To account for this, we transform the stability diagram linearly for the diamond edges to be symmetric by the bias window before calculating the logarithm of the ratio of the currents at positive and negative bias (Fig. \ref{fig3}b). We extrapolate the current between the experimental points, and set a cut-off at the noise level.

Figure \ref{fig3}c shows cuts of the experimental data at the resonance points of the $0\leftrightarrow 1$ excess charge transition, the line connecting the $E(N)=E(N+1)=E(N+2)$ degeneracy points, and the $1\leftrightarrow 2$ transition, depicted as dashed lines in corresponding colours in Fig. \ref{fig3}a, and as thick solid lines in panel c. For the $0\leftrightarrow 1$ and $1\leftrightarrow 2$ transitions, the logarithm of the current ratio is close to $\ln 4$ and $0$ in the large-bias single-transfer regime (upper and lower panels in Fig. \ref{fig3}c), giving the entropy differences $\Delta S_{01}=\pm \ln 4$, $\Delta S_{12}=0$. The deviation from the Hubbard model (dashed grey lines), which is discussed in detail below, at higher bias (shaded areas) is due to vibrational effects, as outlined in Appendix \ref{vibrations-sec}. On the middle plot, the line cuts through a region with two energetically accessible charge state transitions, the value of the current ratio logarithm approaches $\ln 2$ at high bias. Using Equation \ref{eq-two} and the two entropy differences found from the single-process charge transitions, we find $d_{10}/d_{12}=1/2$. The entire line of the cut (orange in Figs \ref{fig3}a,b) is on resonance and there is little deviation from the theoretical curve. 

\subsection{Thermodynamic deduction and comparison to the Hubbard model}
\begin{figure}
\centering
\includegraphics[width=0.49\textwidth]{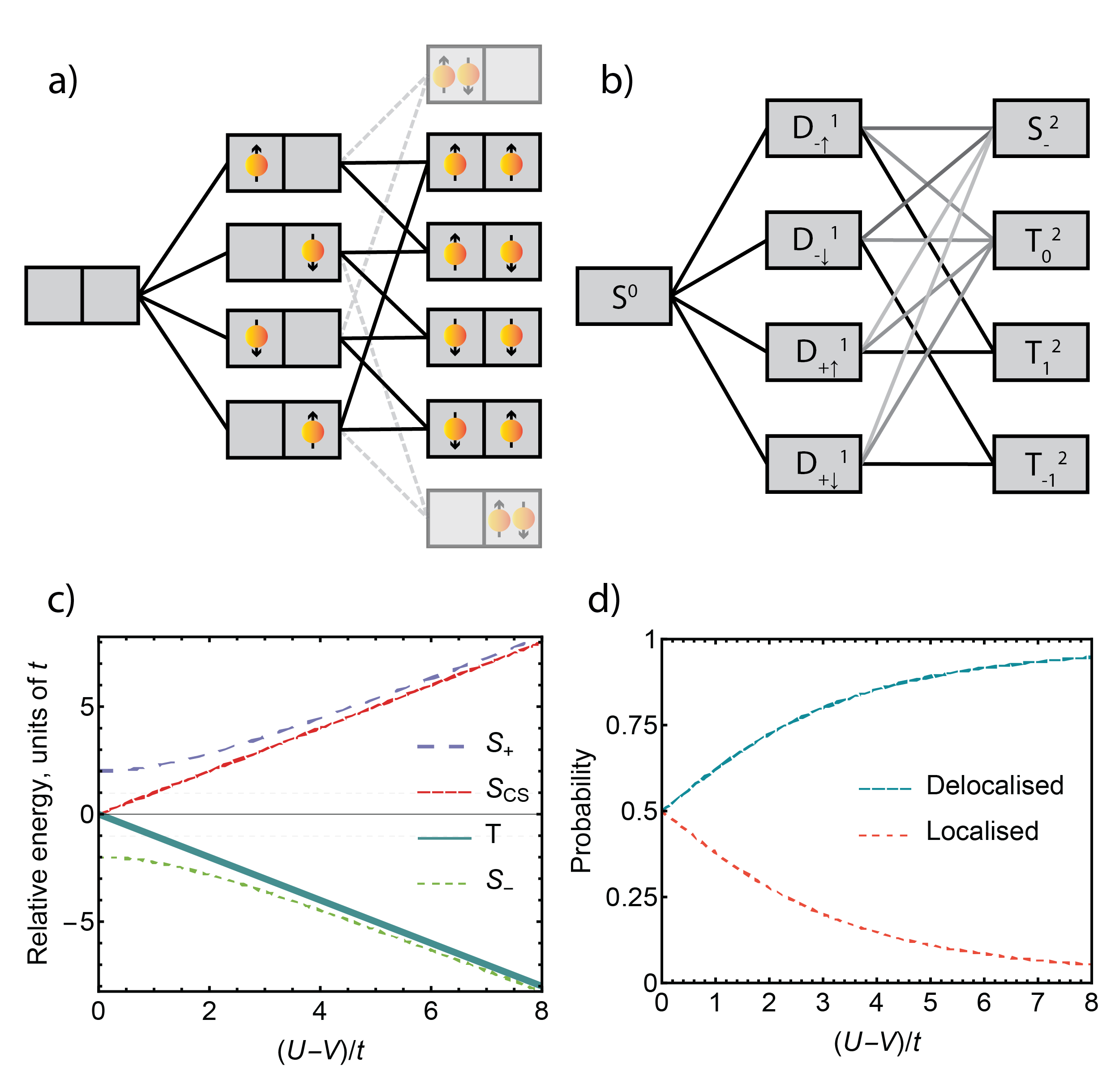}
\caption{a) Classical electron configurations for each of the charge-states of a dimer. The selection rules are indicated by the connecting lines. The two high-energy localised states that do not contribute to transport are shaded. b) Selection rules for the quantum electron states contributing to conductance. The intensity of the colour of the edge of the graph represents the Dyson coefficient for the transition. c)Relative energies of the Hubbard energy levels (shifted by $(U+V)/2t$ to achieve a mean of zero) of the $N=2$ charge-state as a function of the delocalisation parameter $(U-V)/t$. d) The probabilities of finding the molecule in the $S^2_-$ ground state in the delocalised (electrons occupy different spacial sites) and localised (electrons occupy the same site) states as a function of $(U-V)/t$.}
\label{fig4}
\end{figure}

We can use the results of the current ratio analysis to deduce the microscopic dynamics of the molecule in question, even with no knowledge of its structure. The entropy changes of $\kb \ln 4$ and $0$ from Fig. \ref{fig3}c for the $N\leftrightarrow N+1$ and $N+1 \leftrightarrow N+2$ transitions respectively indicate either $1,4,4$ or $4, 1, 1$ (or multiples of these) for the microstate multiplicities of the charge states in the order of excess charge. For simplicity, we choose the first option, as a singlet-to-singlet transition with the addition of an electron is difficult to explain. 

A transition from a single microstate to four between $N$ and $N+1$ charge-states suggests a two-fold spatial degeneracy in addition to the two-fold spin degeneracy, in order words, a dimer. Figure \ref{fig4}a shows a classical graph for the microstates of electrons occupying a two-site system and the allowed transitions between them. In the $N+2$ charge state six classical microstates are expected, in contrast to four seen in the experimental data. This discrepancy can be explained if intra-site potential energy (i.e. the Hubbard parameter $U$) is large enough such that the two states with both electrons occupying the same site are high in energy, outside of the experimentally applied bias window, and do not contribute to charge transport. Importantly, with the two localised states removed, the graph (Fig. \ref{fig4}a) also holds the experimentally derived property of $d_{10}/d_{12}=1/2$ -- every microstate in the $N+1$ state has one $N$ microstate available to it, but two (out of four) in $N+2$, matching the data in Fig. \ref{fig3}c (middle panel) . 

Whilst this simple deductive process cannot be conclusive, it does allow us to infer the structure of the system and rule out potentially many options that would give the same signatures in the experimental data maps. Notably though, it does not take quantum effects into account. In order to see the efficacy of our proposed classical model, and the extent of ``quantumness'' necessary to describe the molecular dynamics, we compare it to the Hubbard Hamiltonian approach given in \cite{Thomas2021} that models the transport through analysis of the molecular orbitals.

Two site orbitals, localised on the (electron-rich) anchoring groups at either end of the molecule, can be generated from taking linear combinations of the two highest occupied molecular orbitals\cite{Thomas2021}. The electronic structure of the molecule is calculated in the charge states that result from the occupation of these site orbitals, i.e. from $N$ when the two sites are empty (corresponding to the molecule in the +4 oxidation state), to $N+4$ when both sites are doubly occupied, which corresponds to a neutral molecule, from the extended Hubbard dimer Hamiltonian. This gives the energies of the microstates involved in electron transfer in terms of the intra-site, $U$, and inter-site, $V$, potential terms, and the kinetic term, $t$, (see Figure \ref{fig1}c and Table \ref{table-states}) as well as the Dyson coefficients, $D_{jk} = \bra{\phi_{k}}a_{i, \sigma}^{+}\ket{\phi_{j}}$, that encode the selection rules for electron transfer between two many-body quantum states: $\phi_{j}$ and $\phi_{k}$. $a^{+}_{i, \sigma}$ is the creation operator for an electron of spin $\sigma$, in the site orbital $i$\cite{Fransson06}.

The values of current ratios found from fitting the full experimental stability diagram to the Hubbard model ($U$, $V$ and $t$ are 0.5, 0.14 and 0.01 eV respectively)\cite{Thomas2021}. The results predicted by this fitted model are shown as dashed grey lines on Fig. \ref{fig3}c. 

The Hubbard parameters can also be used to calculate the degree of electron localisation -- the coefficients $c_{+/-}$ describing the wavefunctions (see Table \ref{table-states}) depend on $U$, $V$ and $t$ \cite{Thomas2021}:
\begin{equation}
\label{eq-cs}
    c_{+/-}=\frac{1}{2}\sqrt{1\pm \frac{U-V}{2C}}
\end{equation}
where 
\begin{equation}
    C=\sqrt{\left(\frac{U-V}{2}\right)^2+4t^2}
\end{equation}

Figure \ref{fig4}c shows the energy dependence of the $N=2$ states in units of $t$  as a function of the main dimensionless parameter of the model: $(U-V)/t$. Figure \ref{fig4}d shows the probability of measuring the ground state $S^2_-$ in the localised ($\ket{\uparrow \downarrow; 0}$ or $\ket{0;\uparrow \downarrow}$) state vs. the delocalised ($\ket{\uparrow; \downarrow}$ or $\ket{\downarrow; \uparrow}$).  

A large value of $(U-V)/t$, as given by the Hubbard model fitting \cite{Thomas2021}, results in: (i) the singlet ground state ($S_-^{2}$) being open shell and energetically close to the triplet ($T^{2}$), (ii) the high energy of the two remaining $N=2$ singlet states with significant closed-shell character compared to $S_-^{2}$ and $T^{2}$, and (iii) the relatively small energy splitting between two $N+1$ doublet states $D^{N+1}_{+}$, $D^{N+1}_{-}$ (the dashed grey lines in Fig. \ref{fig4}c indicate the $2t$ splitting between the doublets).

The graph in Figure \ref{fig4}b shows the quantum states involved in charge transport and has a similar structure to the classical graph (Fig. \ref{fig4}a), but with the lines connecting the states weighted by the Dyson coefficients, dependent on $U$, $V$, and $t$ \cite{Thomas2021}, -- the energy-inpedendent contribution to the transfer probability ($d_{nn'}$) is no longer 0 or 1, indicated by the presence or absence of an edge between the microstates in the classical graph, but is given by the Dyson coefficient of the transition.  

The effects of quantum correlation on transport, captured within the Hubbard model and derived from the molecular orbital structure, match well with the the classical system-agnostic fluctuation-relation approach. The reason behind this is that, while the eigenstates of the Hubbard Hamiltonian presented in the site basis (Figure \ref{fig4}b and Table \ref{table-states}) are linear combinations of the classical states (Figure \ref{fig4}a), the degree of delocalisation does not change the overall number of microstates. 

Furthermore, $d_{10}/d_{12}=1/2$ is a fundamental value, as half of the phase-space in the $N+2$ state will be inaccessible for any given spin orientation in the $N+1$ state. This result is graphically shown in Fig. \ref{fig4}a,b) and has been confirmed to be parameter-independent by an explicit calculation of Dyson coefficients (see Appendix \ref{dyson-sec}). 

Different behaviour would be observed for a system where $(U-V)/t$ is small. In this case, the ground state may be the only one accessible within the bias window (see Fig. \ref{fig4}c), the energy splitting between the doublet states in the $N=1$ state is large and the system essentially behaves as a single-site quantum dot. This leads to current ratios of 1-2-1 in the single-transfer regime, which has been observed experimentally in charge transport measurements of shorter porphyrin monomer devices\cite{Limburg2019}.

In addition to a general description of the molecule as a dimer with a high degree of delocalisation, we are able to determine a lower bound on $(U-V)/t$. The fact that we are not able to resolve the two doublets of the $N=1$ state in the top panel of Fig \ref{fig3}c means that $2t<\kb T$, for the experimental temperature of 77 K. At the same time, in the lower panel of Fig \ref{fig3}c at the highest single-rate bias of approximately 0.3V, we do to reach the higher-lying singlet states. Combined together, these two facts lead to $(U-V)/t>5$, or, otherwise, the probability of finding the ground state to be delocalised is above 0.67, which is a measure of correlation within the system. 
\section{Discussion and conclusions}
In this work, we presented a fully-thermodynamic fluctuation-relation current analysis method. The proposed analysis method is applicable to a common experimental setup -- out-of-equilibrium charge transport through a single-electron device. The main feature distinguishing our analysis protocol from previous thermodynamic methods, focused on measuring entropy in equilibrium or quasi-equilibrium states, is its applicability to highly out-of-equilibrium conditions. This introduces a new energy scale and allows for high-energy microstates to be uncovered, ones that would not normally be resolved by equilibrium methods due to their infinitesimal populations and contributions to current.

We carried out a proof-of-principle test of the method on a single-molecule device described by the Hubbard dimer model, a ubiquitous description of correlated electron systems, and showed that it is possible to make deductions about the electronic structure of the molecule, including the degree of electron correlation and entanglement\cite{salfi16} in the ground state of the system and excited states of this system. This demonstrates promise for using the method to unravel the microscopic details of more exotic correlated electron states.

A strength of the method is that it is based on the non-equilibrium fluctuation theorem, a general stochastic thermodynamics result. We believe that the analysis of few-electron nanodevices from the stochastic thermodynamics point of view  is a promising direction, as, while systems open not only to energy, but also to particle exchange, are not the most frequent object of study in the field, they offer a broad and technically well-developed experimental platform.

\section*{Acknowledgements}
J.A.M. was supported through the UKRI Future Leaders Fellowship, Grant No. MR/S032541/1, with in-kind support from the Royal Academy of Engineering. J.O.T was supported by the EPSRC (Grant No. EP/N017188/1).
\\

\appendix
\section{Mean system population}
\label{proof-sec}
While the non-equilibrium fluctuation relation is general, it is not frequently applied to electric nanodevices. As further proof of the applicability of the fluctuation relation approach, we find the mean additional population of the system in the single-transfer regime $n=\left< N \right>-N$, where  $\left< N \right>$ is the time-averaged population, and the values of $n$ are between 0 and 1.

The rate equation \cite{Nazarov2009, Harzheim2020} gives $n$ as $n=\Gamma_T/(\Gamma_F+\Gamma_T)$. Using the ratio of $\Gamma_T/\Gamma_F$ from the non-equilibrium fluctuation relation (Eq. \ref{eq-theorem}), we find:
\begin{equation}
    n=\frac{1}{1+e^{\frac{\varepsilon-T\Delta S}{\kb T}}}
\end{equation}
-- a Fermi-distribution shifted by $T\Delta S$ in energy. This is in  agreement with the result in \cite{Pyurbeeva2020b}, which was derived both from fully thermodynamic considerations and from the Gibbs distribution for a system with arbitrary electronic energy levels. 

\section{The role of vibrations}
\label{vibrations-sec}
\begin{figure}
\centering
\includegraphics[width=0.5\textwidth]{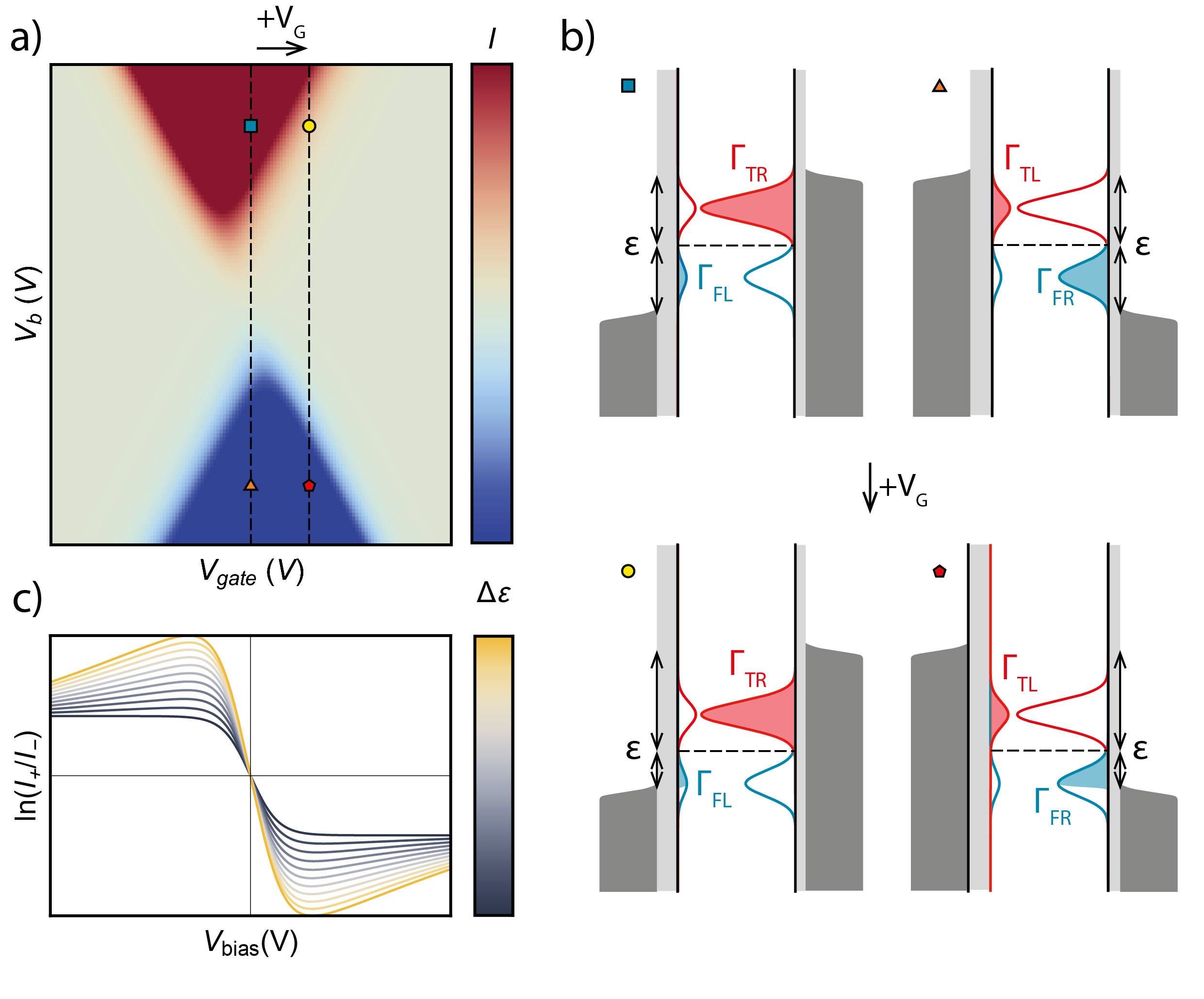}
\caption{a) A sample modelled stability diagram in the presence of electron-vibrational coupling, $d_{01} = 2$, $d_{10} = 1$, and $\gamma_R/\gamma_L = 5$. It can be seen that the IV-trace is symmetric by bias voltage at the resonance point and asymmetric off-resonance. b) Energy diagrams of electron transfer processes in the presence of vibrational coupling on- (above) and off- (below) resonance. The diagrams correspond to the points on the stability diagrams marked by the same symbol. c) The effect of detuning from the resonance on the logarithm of the modelled current ratios.}
\label{fig5}
\end{figure}
The theoretical approach taken above does not include vibrational effects. From a thermodynamic viewpoint, energy loss to the phonon bath is associated with an additional entropy change, which leads to a discrepancy in the entropy differences associated with an electron hopping in and out of the system, thus yielding the non-equilibrium theorem inapplicable to the system. 

While the extension of the non-equilibrium fluctuation theorem to finding relative probabilities of fluctuations associated with entropy changes of different values is of large scientific interest, here we apply the more standard rate equation approach. In the weak molecule-electrode coupling regime \cite{Seldenthuis08, Thomas2019}, electron-vibration coupling introduces an energy-dependence $k(\varepsilon)$  to the hopping rates in equations \ref{eq-rates}:
\begin{equation}
    \begin{dcases}
    \label{eq-vib}
    \Gamma_{T L/R}=\gamma_{L/R} d_{01}\int f_{L/R} k(\varepsilon)d\varepsilon
    \\~\\
    \Gamma_{F L/R}=\gamma_{L/R} d_{10}\int \left( 1-f_{L/R} \right) k(-\varepsilon)d\varepsilon
    \end{dcases}
\end{equation}

The energy-dependence of the electron-vibration coupling is symmetric around $\varepsilon$, therefore the approach outlined above is unaffected on resonance as the bias window opens symmetrically around $\varepsilon$ and the $k(\varepsilon)$ contributions cancel. 

However, with asymmetric tunnel coupling, the electron-vibration coupling results in an asymmetry of the sequential tunnelling region in gate and bias voltage with respect to the resonance point \cite{Limburg2019}. This effect is demonstrated in Fig. \ref{fig5}a, which displays a stability diagram calculated with a simple Marcus Theory approach to electron-vibration coupling to obtain $k(\varepsilon)$ as thermally broadened Gaussians, $d_{01} = 2$, $d_{10} = 1$, and $\gamma_R > \gamma_L$ \cite{Sowa18}. Fig. \ref{fig5}b show that on resonance (top two panels), the energy-dependent term cancels at each $V_{bias}$ when taking the current ratios, which is not the case for off-resonance situations (bottom two panels). Fig \ref{fig5}c, shows that as the resonance is detuned (grey to yellow), the bias voltage at which the logarithm of the current ratio tends to the expected value of $\ln 2$ increases as, off-resonance, both energy-dependent functions must been integrated for the terms to cancel in \ref{eq-vib}. This explains why electron-vibration coupling causes the entropic analysis to deviate when studying off-resonance current ratios. This regime came into play as the line cuts in Fig. \ref{fig3}c moved from a single transition area to a two-transition one, defined by the grey shaded areas. 

\section{Dyson coefficients}
\label{dyson-sec}
\begin{table}
\caption{\label{dyson-tab}The Dyson coefficients for the transitions between the N=1 and N=2 charge-states.}
\begin{ruledtabular}
\begin{tabular}{lllllll}
&\makecell{$S^2_-$} & \makecell{$T^2_1$} &  \makecell{$T^2_0$} &  \makecell{$T^2_{-1}$}&\makecell{$S^2_{CS}$}&\makecell{$S^2_{+}$}\\
\hline\\
$D^1_{+,\uparrow}$ & $\dfrac{(c_++c_-)^2}{2}$ & $\dfrac{1}{2}$ & $\dfrac{1}{4}$ & 0 & $\dfrac{1}{4}$ & $\dfrac{(c_+-c_-)^2}{2}$
\\[10pt] 
$D^1_{+,\downarrow}$  & $\dfrac{(c_++c_-)^2}{2}$ & 0 & $\dfrac{1}{4}$ & $\dfrac{1}{2}$ & $\dfrac{1}{4}$ & $\dfrac{(c_+-c_-)^2}{2}$
\\[10pt] 
$D^1_{-,\uparrow}$ & $\dfrac{(c_+-c_-)^2}{2}$ & $\dfrac{1}{2}$ & $\dfrac{1}{4}$ & 0 & $\dfrac{1}{4}$ & $\dfrac{(c_++c_-)^2}{2}$
\\[10pt] 
$D^1_{-,\downarrow}$ & $\dfrac{(c_+-c_-)^2}{2}$ & 0 & $\dfrac{1}{4}$ & $\dfrac{1}{2}$ & $\dfrac{1}{4}$ & $\dfrac{(c_++c_-)^2}{2}$
\\[5pt]
\end{tabular}
\end{ruledtabular}
\end{table}

Table \ref{dyson-tab} shows the Dyson coefficients for the transitions between all electronic levels from the N=1 to the N=2 charge-states.

In our molecular device, all four states in the $N=1$ charge-state are equally occupied (as thermal broadening prevents the observation of the $\ln2$ shoulder on the top panel in Fig. \ref{fig3}c). This means that the total probability of transition to each of the electronic levels in the $N=2$ charge-state is proportional to the sum over the corresponding column in Table \ref{dyson-tab}. 

It can be shown that $(c_++c_-)^2+(c_+-c_-)^2=1$ (see Equation \ref{eq-cs}), and thus the sum in each column, and the total energy-independent contributions to the transition probability to each of the levels in the $N=2$ charge-state are equal. Thus, the selection rules term $d_{12}/d_{10}$ will be equal to half the number of electronic levels in the $N=2$ charge state involved in charge transport, independent of this number. 

This, however, would not have been the case if the two pairs of doublet states $D^{1}_{-}$ and $D^{1}_{+}$ could be resolved. Then, the energy-independent part of the transition probability to $S^2_-$ would differ from one, which would allow to find the values of $c_-$ and $c_+$, not just the boundary.

\section{Selection rules in the single transfer regime}
While we have discussed the effects of selection rules in the two-charge-state-transition regime, we have never mentioned them in the single-transition case. Here, we show that in this case they do not play a role, as long as the microstates of both charge-states form a connected graph -- if any microstate can be reached from any microstate in a finite number of transitions.   

The physical meaning of the system-dependent rate coefficient $d_{01}$ is the mean volume of the phase space in the $N+1$ macrostate the system can occupy if it has transferred to it from a single microstate of the $N$ macrostate, where the mean is taken over the microstates of the $N$ charge-state. If transitions between all microstates of $N$ and $N+1$ are allowed and have the same Dyson coefficient, equal to 1, this volume is the same for each microstate in $N$ and equal to $\Omega_{1}$ -- the number of microstates in $N=1$. However, in the presence of selection rules, it is not the case.

In the single-transfer areas, by equation \ref{eq-ratio}:
\begin{equation}
\label{rules}
    \frac{d_{01}}{d_{10}}=\frac{D_{01}}{\Omega_0} \frac{\Omega_1}{D_{10}}
\end{equation}
where $D_{01}$ is the sum of all Dyson coefficients for the transitions from $N$ to $N+1$ and $\Omega_0$ is the number of microstates in $N$. In the classical case of a graph, $D_{01}$ is the number of edges leading from all the microstates in $N$ towards $N+1$. Since each independent transition is equally likely to happen in both directions, due to the Fermi golden rule, every edge has two ends, the ratio ${d_{01}}/{d_{10}}$ is simply the ratio of the microstate numbers in the charge states.


\bibliography{refs}

\end{document}